# Inducing ferroelectricity in $NH_4I$ and $NH_4Br$ via partial replacement of protons by deuterons

*Miao Miao Zhao,[1‡] Lei Meng,[1‡] Yi Yang Xu,[1] Na Du,[1] Fei Yen[1*]*

[1]School of Science, Harbin Institute of Technology, Shenzhen, University Town, Shenzhen, Guangdong 518055, P. R. China

ABSTRACT. While all of the polymorphs of $NH_4I$ and $NH_4Br$ are non-polar, a reversible electric polarization is established in the ordered γ phases of $(NH_4)_{0.73}(ND_4)_{0.27}I$ and $(NH_4)_{0.84}(ND_4)_{0.16}Br$ (where D is $^2H$) via *dc* electric fields. The presence of two groups of orbital magnetic moments appears to be responsible for the asymmetric lattice distortions. Our findings provide an alternative pathway for hydrogen-based materials to potentially add a ferroelectric functionality.

INTRODUCTION

At room temperature, the ammonium ions ($NH_4^+$) in most solid compounds continue to exhibit motions in the form of reorientations. Extremely weak magnetic moments $\mu_p$ are generated when $NH_4^+$ ions reorient by their most elemental steps of 120° ($C_3$) or 90° ($C_4$) since a simultaneous hopping of 3 or 4 protons ($H^+$) trace out complete current loops (3 protons covering 120° or 4 protons each covering 90° of a circle).[1,2] The electrons reside near the N atom so they mainly yield a diamagnetic contribution that is independent of temperature. Given the tetrahedral symmetry of the $NH_4^+$, there are 8 possible types of $C_3$ reorientations and 6 types of $C_4$ reorientations, each having an associated $\mu_p$. Above a certain critical temperature $T_C$, all 14 reorientations are energetically accessible, meaning that the magnetic moments of $NH_4^+$ ions behave 'paramagnetically' but there are only 14 possible directions for $\mu_p$ to point to. Below $T_C$, the more energetically costly $C_4$ reorientations seize.[3] The remaining 8 internal degrees of freedom cause the system of $NH_4^+$ to become geometrically frustrated because a fraction of neighboring $NH_4^+$ end up having their $\mu_p$ pointing along the same direction. This causes a problem because neighboring protons having similar orbitals generate resonant forces so the system phase transitions into a slightly augmented lattice as a response where the $\mu_p$ become modulated. Taking ammonium sulfate as an example, the $SO_4^{2-}$ ions slightly shift in the ferroelectric phase. However, since there are two groups of $NH_4^+$, two types of dipole moments are formed with the $SO_4^{2-}$ that cannot completely cancel out so a net electric polarization emerges.[4] This perspective can be applied to further explain why substitution of alkali metals by $NH_4^+$ in some inorganic compounds[5,6] as well as a presence of two or more sets of proton orbitals in some molecular frameworks can produce a ferroelectric phase.[7-9] An interesting question now is, can we modify compounds that only have one type of $NH_4^+$ into having two types in the attempts to generate ferroelectricity?

One of the simplest ammonium-based compounds with only one type of $NH_4^+$ are the ammonium halides $NH_4X$ (where X = Cl, Br and I). At room temperature, these compounds have either a NaCl-type or CsCl-type of cubic structure where the N atom is surrounded by 8 equidistant $X^-$ ions with the $H^+$ nearly midway in between.[10] This gives rise to two possible spatial configurations, *A* and *B*, shown in Figure 1a. The $C_3$ and $C_4$ reorientations are now easier to picture as $NH_4^+$ reorienting about the 8 diagonals and 6 axes perpendicular to the faces of a cube, respectively, by 120° and 90°.[11-13] Below $T_C$, the $NH_4^+$ become ordered into an *ABAB...* checkerboard pattern along the *ab*-plane[10] from which we presume is a manifestation of the $\mu_p$ becoming ordered in such a way as to trace out a spiral modulation (along the four main diagonals of one side of the face of the cube) to minimize internal resonance[14] since the $C_4$ and $C_2$ reorientations are no longer active.[3] The $X^-$ ions also end up off-centering along the *c*-axis in a modulated fashion[10] rendering the low temperature γ phase antiferroelectric.[15] Suppose now that a fraction of the $\mu_p$ was to be replaced by a second type of magnetic moment with a different magnitude $\mu_D$, so that the off-centering of the $X^-$ ions cannot fully cancel out (Figure 2b), then a ferrielectric phase may be induced such as in the case of ammonium sulfate. Technically, this phase should be ferrielectric because the directions of the electric dipole moments *p* do not all point along the same direction, instead, the antiparallel configuration of neighboring *p* are only tailored to lose symmetry. This substitution may be attainable by replacing a fraction of the $NH_4^+$ by $ND_4^+$ because the latter ions reorient at a slower rate due to the more massive $D^+$ ions. A ratio of 1:1 or one that is commensurate to the correlation length may not be enough to render the superstructure to become non-centrosymmetric since the two modulations will have a tendency to couple in order to maintain the antiferroelectric environment; a sort of electric analogue of Lenz' Law where the system is simply trying to preserve the original crystal field. However, employing an incommensurate ratio of $H^+$ and $D^+$ in the systems $(NH_4)_x(ND_4)_{1-x}I$ and $(NH_4)_x(ND_4)_{1-x}Br$ or even $NH_4I_{1-x}Br_x$ is likely to circumvent this response because the modulations cannot completely cancel each other out.

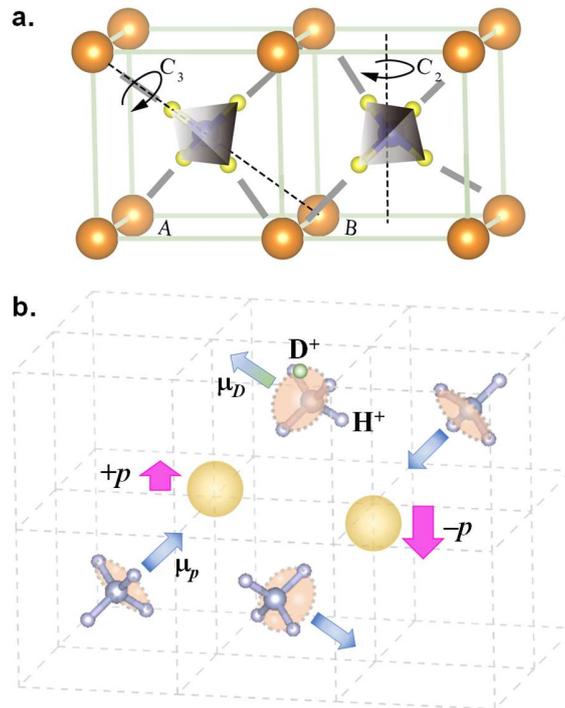

**Figure 1.** (a) Possible types of reorientations that $NH_4^+$ tetrahedra exhibit in $NH_4I$. Either a $C_3$ (8) or $C_2$ (6) reorientation by 120° or 180° about a face or edge of the tetrahedron preserves the symmetry of the lattice. A $C_4$ reorientation by 90° is also possible, but the spatial configuration will change from *A* to *B* or vice versa. The iodine atoms are located at the vertices of the cubes. Dark grey lines represent hydrogen bonds. (b) In the ordered γ phase, the magnetic moments (blue arrows) of the protons ($H^+$) $\mu_p$ associated to the $NH_4^+$ reorientations order and the iodine atoms shift off-center along the *c*-axis generating alternating dipole moments *p* (pink arrows). By partially substituting $H^+$ by $D^+$, a second type of magnetic moment $\mu_D$ may cause $\Sigma\, p \neq 0$ and render the system non-polar.

In this work we show that the electric polarizations of the ordered γ phases in $(NH_4)_{0.73}(ND_4)_{0.27}I$ and $(NH_4)_{0.84}(ND_4)_{0.16}Br$ can be switched by *dc* electric fields above $T_C$.

Magnetic susceptibility measurements were performed to support the idea that the structural phase transitions are magnetically-driven. Dielectric constant measurements were also carried out to verify the quality of the crystals and discern the well-known hysteretic nature of the ordered γ phase and disordered α and β phases of the ammonium halide family. Our findings open up the possibility of hydride compounds that are already magnetic to potentially becoming multiferroic via partial isotope exchange.

EXPERIMENTAL METHODS

Partially deuterated ammonium iodide crystals $(NH_4)_{0.73}(ND_4)_{0.27}I$ were grown from slow evaporation of a solution of $NH_4I$ 99.999% in purity (CAS No. 12027-06-4) dissolved in a mixture of deionized $H_2O$ and 99.5% deuterated heavy water (CAS No. 7789-20-0) in a molar ratio of 2:1. The same process was employed to grow crystals of $(NH_4)_{0.84}(ND_4)_{0.16}Br$ but using $NH_4Br$ 99.99% in purity (CAS No. 12124-97-9) dissolved in $H_2O$ and $D_2O$ mixed with a ratio of 5:1. All reagents were purchased from Aladdin-e, Inc. Shanghai. For the $(NH_4)_{0.84}(ND_4)_{0.16}Br$ case, urea was added as a catalyst to avoid formation of dendritic structures.[16] All samples used were transparent, cuboid in shape and at least 2 x 2 x 2 mm$^3$ in size. To determine the doping concentrations, a fraction of the crystals was grounded back into powdered form, pressed into pellets and the magnetic susceptibility vs. applied field curves measured at 300 K. The obtained magnetic susceptibility was then matched to a linear fit (Vegard's rule) of the magnetic susceptibilities of the undoped (i.e. fully hydrogenated and fully deuterated) samples. The structures were drawn with the aid of the VESTA (Visualization of Electronic and STructural Analysis, v. 3.4.7) freeware.[17]

The magnetic susceptibility was measured with the VSM option of a PPMS (Physical Properties Measurement System) fabricated by Quantum Design, Inc. San Diego. Samples typically of at least 8 mm$^3$ were used. The samples were attached onto the manufacturer's standard quartz tube sample holder with GE varnish. The contribution of the varnish was corrected. The pyroelectric current was measured by a Keithley 6517B electrometer during warming. The temperature was controlled by the cryostat of the PPMS. The samples were grinded to less than 0.5 mm thick and sandwiched between two parallel electrode plates painted with silver epoxy to form a capacitor-like device. An electric field was applied to the sample at room temperature followed by cooling to a base temperature. Then, the electric field was removed and the two electrode contacts shorted for 30 minutes to remove residual charge build-up. The sample was then warmed back up to room temperature at a rate of 2 K/min while measuring its current with the electrometer. For the measurements of the real and imaginary parts of the dielectric constant, the same type of capacitor-like samples was used. An Agilent E4980AL LCR impedance meter measured the capacitance and loss tangent of the sample.

RESULTS AND DISCUSSION

Upon cooling, pure NH$_4$I phase transitions from a NaCl $\alpha$ phase to a CsCl $\beta$ phase at $T_{\alpha-\beta\_I}$ = 256 K and to a tetragonal-like $\gamma$ phase at $T_{\beta-\gamma\_I}$ = 232 K.[10,18] The $A$ and $B$ configurations of the NH$_4^+$ (Figure 1a) in the $\alpha$ and $\beta$ phases are disordered while the ordered $ABAB$ checkerboard pattern emerges in the $\gamma$ phase.[10] At $T_{\beta-\gamma\_I}$, the magnetic susceptibility $\chi(T)$ with respect to temperature was found to increase by ~5% due to the long-range ordering of the proton orbitals.[1] Upon warming, the $\gamma$ phase remained stable up to $T_{\gamma-\alpha}$ = 281 K where the system transformed back to the $\alpha$ phase. The $\alpha$, $\beta$ and $\gamma$ phases of the ammonium halides have been reported by many researchers to be capable of being supercooled, superheated and coexist in very large temperature ranges.[3,13,18-24]

Figure 2a shows $\chi(T)$ of the partially-deuterated crystal $(NH_4)_{0.73}(ND_4)_{0.27}I$ measured along the *b*-axis direction under an applied magnetic field of $H = 1$ T. The β phase appeared to reversibly phase transition into and out of the γ phase at the transition temperatures of $T_{\beta-\gamma\_73:27}$ = 242 K and $T_{\gamma-\beta\_73:27}$ = 262 K. However, upon further warming past 262 K, $\chi(T)$ exhibited a minimum in the temperature region between $T_{\gamma-\beta\_73:27}$ and $T_{\beta-\alpha\_73:27}$ = 282 K which we interpret to be a coexistence of the α, β and γ phases due to the following two reasons. First, the γ phase was observed that it can coexist with the α and β phases up to 293 K according to two independent sets of proton spin-lattice relaxation experiments.[3,13] Second, according to our pyroelectric current measurements presented below, the non-zero electric polarization, which is associated to the ordered γ phase, vanishes exactly at $T_{\beta-\alpha\_73:27}$ = 282 K.

Figure 2b displays the electric polarization of $(NH_4)_{0.73}(ND_4)_{0.27}I$ obtained from integrating the area below the measured pyroelectric current curves (shown in the inset of Figure 2b). To polarize the sample, an electric field of $E = 6$ kV/cm was applied at 300 K and the sample was cooled down past $T_{\beta-\gamma\_73:27}$ to 230 K. The electric field was then removed and the pyroelectric current was measured during warming. The polarity of the pyroelectric current was switchable according to the polarity of $E$ which is evidence that the ordered γ phase of $(NH_4)_{0.73}(ND_4)_{0.27}I$ is ferroelectric. Both the positive and negative pyroelectric currents reverted to zero nearly exactly at $T_{\beta-\alpha\_73:27}$, which agrees with the notion that the ordered γ phase persisted up to this temperature.

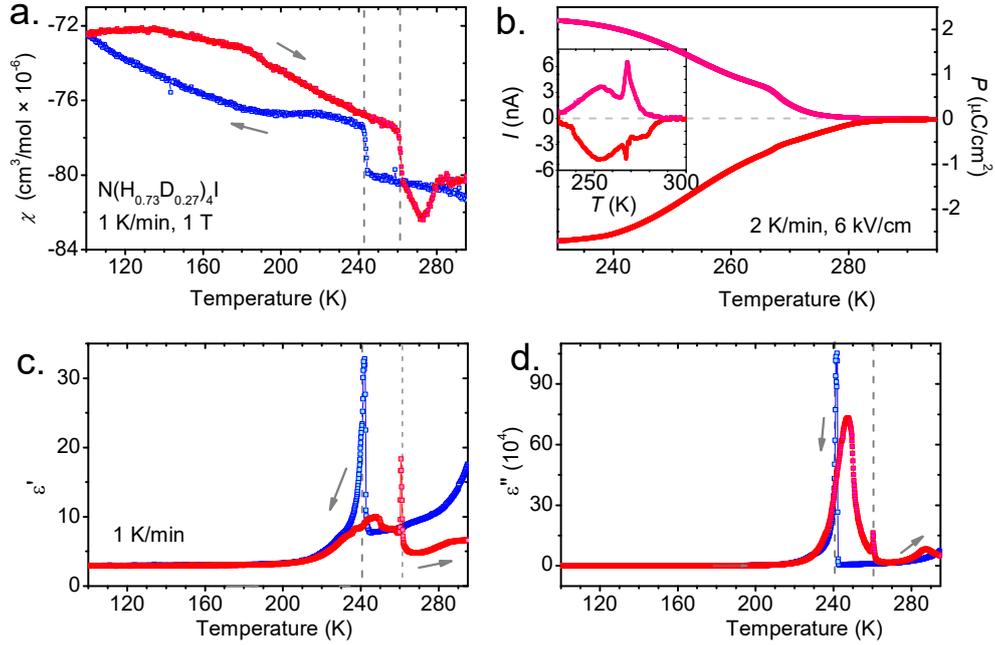

**Figure 2.** (a) Molar magnetic susceptibility $\chi(T)$ of $(NH_4)_{0.73}(ND_4)_{0.27}I$ under an applied magnetic field of $H = 1$ T along the *b*-axis. Arrows indicate the cooling and warming curves; vertical dotted lines indicate phase transition boundaries. (b) Electric polarization $P(T)$ obtained from integration of the measured pyroelectric current during warming (inset) under a bias electric field of $E = 6$ kV/cm. (c) Real part $\varepsilon'(T)$ and (d) imaginary part of the dielectric constant $\varepsilon''(T)$ measured at 1 kHz and 20 V/cm.

Figures 2c and 2d show the real $\varepsilon'(T)$ and imaginary $\varepsilon''(T)$ parts of the dielectric constant measured at 1 kHz. Sharp peaks were observed at 241 K during cooling, in excellent agreement with the discontinuity at the same temperature in $\chi(T)$ reflecting the β to γ phase transition. Upon warming, the coinciding sharp peaks were observed at 261 K, also in good agreement with the associated discontinuity in $\chi(T)$. However, the peaks during warming were smaller in magnitude, an indication that the phase transition out of the γ phase was only partial at this temperature which explains why the electric polarization (Figure 2b) persisted up to near 282 K.

The α to β and β to γ transition temperatures of pure $NH_4Br$ occur at $T_{\alpha-\beta\_Br} = 411$ K and $T_{\beta-\gamma\_Br} = 235$ K, respectively.[25,26] The α, β and γ phases of $NH_4I$ and $NH_4Br$ are respectively isostructural to each other so below $T_{\beta-\gamma\_Br}$ the configurations of the $NH_4^+$ should also become ordered for the bromide congener.[10] According to the magnetic susceptibility of $(NH_4)_{0.84}(ND_4)_{0.16}Br$ (Figure 3a), the disorder-order transition temperature shifted to $T_{\beta-\gamma\_84:16}$ = 247 K upon 16% deuteration. The associated sharp step-up anomaly by nearly 2% at $T_{\beta-\gamma\_84:16}$ indicates that a) the phase transition is first order in nature and b) the sample is single-phased, i.e. the deuterons are statistically distributed throughout the lattice. During warming, the coinciding step-down discontinuity occurred at $T_{\gamma-\beta\_84:16}$ = 262 K, however, above this temperature, $\chi(T)$ continued to decrease to reach a minimum at 282 K in what appears to be the complete transformation of the γ phase to the β phase according to the polarization measurements presented next. Hence, a coexistence of the γ and β phases existed in the temperature range of 262 K and 282 K during warming. These results are in good agreement with a recent finding by Funnel *et al.* where they concluded that the γ phase can remain metastable up to 310 K in $ND_4Br$.[24]

Figure 3b shows multiple runs of the electric polarization of $(NH_4)_{0.84}(ND_4)_{0.16}Br$ under $\pm E$ = 1–1.33 kV/cm. The raw pyroelectric current data is also displayed in Figure 3c along with the polarity and magnitude of the bias voltage for each of the curves. The magnitude of the largest recorded polarization for each curve ranged between 180–810 $nC/cm^2$, a value comparable to that of Rochelle's salt of ~250 $nC/cm^2$ at 278 K.[27] Similar to $(NH_4)_{0.73}(ND_4)_{0.27}I$, the polarization of $(NH_4)_{0.84}(ND_4)_{0.16}Br$ decreased to zero upon warming to 280 K suggesting that the γ phase is no longer present in the system at this temperature. The reversibility of the polarization according to the polarity of $E$ confirms the ferroelectric nature of the ordered γ phase of $(NH_4)_{0.84}(ND_4)_{0.16}Br$.

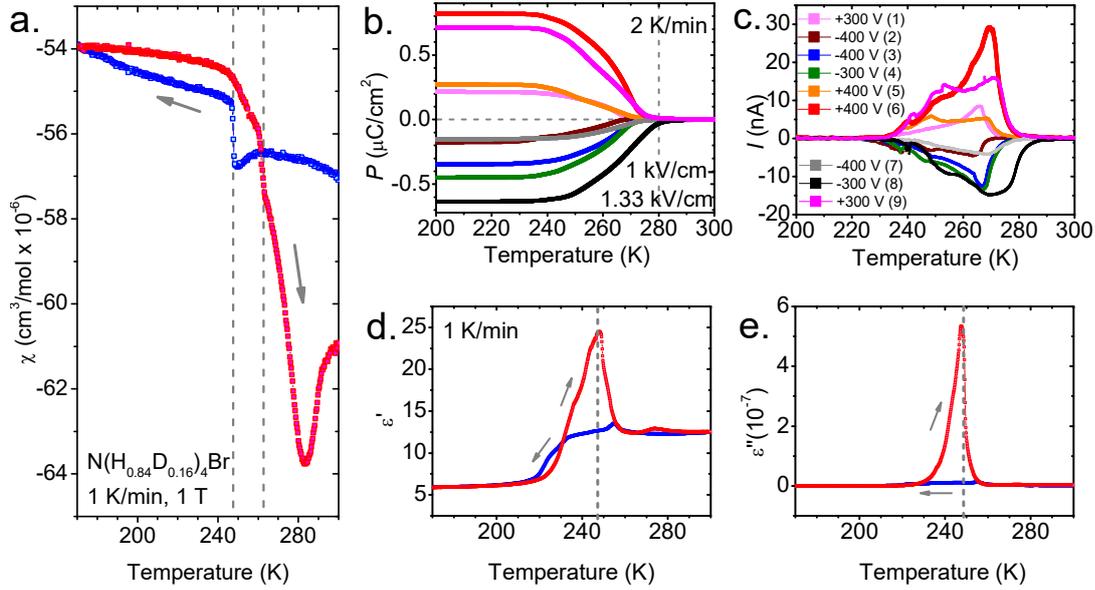

**Figure 3.** (a) $\chi(T)$ of $(NH_4)_{0.84}(ND_4)_{0.16}Br$ under $H = 1$ T. (b) $P(T)$ of a sample subjected to multiple consecutive runs of different polarity and $E$ ranging between 1 and 1.33 kV/cm. (c) Raw data of the measured pyroelectric currents used to obtain $P(T)$. Numbers in parentheses denote the $n$th run. (d) $\varepsilon'(T)$ and (e) $\varepsilon''(T)$ under 1 kHz and 3.3 V/cm.

Figures 3d and 3e show $\varepsilon'(T)$ and $\varepsilon''(T)$ of $(NH_4)_{0.84}(ND_4)_{0.16}Br$. The most pronounced feature are the peaks centered at 248 K during warming, which coincided to $T_{\beta-\gamma\_84:16}$ observed in $\chi(T)$ during cooling. The peaks are rather wide, over 20 K in length, suggestive of the ferroelectric domains not fully disordering upon warming across 248 K. Again, we ascribe to the notion that the system enters into an admixture of the $\gamma$ and $\beta$ phases during warming where a higher order type of phase transition occurs from the $\gamma$ to the $\beta$ phase and completing at 280 K.

In the case of $(NH_4)_{0.73}(ND_4)_{0.27}I$, when 1 out of every 3 protons is replaced by a deuteron, the system will be comprised mostly of $NH_3D^+$ and $NH_4^+$ ions. There will also be some ions in the form of $NH_2D_2^+$, $NHD_3^+$ and $ND_4^+$ present in the system, albeit with lower probabilities. Berret *et al.* found that the correlation length of the antiferroelectric type in $K_{0.38}(ND_4)_{0.62}I$ was 62 Å, or around 10 unit cells.[15] In compounds with full ammonium compositions such as the two samples investigated herein, their correlation lengths are even longer so when the samples enter the ordered γ phase, two main spatial modulations develop that shift the halide ions along the ±$c$-axis by different degrees. Since the $NH_3D^+$ and $NH_4^+$ spiral modulations are incommensurate to each other, the dipole moments of the halide ions do not fully cancel each other out so a ferrielectric phase emerges.

Lastly, the different paths taken by the warming curves in Figure 3c must be addressed. Given the coexistence of the γ and β phases during the warming process, stresses and imperfections arising from lattice constant mismatches are stored in the system. The domain sizes and distributions for each cycle are different and the associated cooperative displacements appear to significantly affect some of the measured macroscopic quantities. To make sure the net polarization in Figures 2b and 3b are not due to transformation stresses, we performed the same experiments on single crystals of the pure compounds $NH_4I$ and $NH_4Br$ and found no spontaneous polarization.

CONCLUSIONS

In conclusion, we identified that ferroelectricity can be established in the ordered γ phases of $(NH_4)_{0.73}(ND_4)_{0.27}I$, and $(NH_4)_{0.84}(ND_4)_{0.16}Br$. The underlying mechanism seems to be due to a presence of two types of orbital-lattice coupling effects that are unable to cancel each other out. Our findings provide a new approach to breaking the spatial-inversion symmetry of superstructures in centrosymmetric space-groups of hydrogen-based compounds. For instance, partial deuteration of materials possessing ions with internal modes such as $NH_4^+$, $CH_3NH_3^+$, $CH_3COO^-$, $BH_4^-$ etc.[28,29] should also enable ferroelectricity to be established in their respective ordered phases.

AUTHOR INFORMATION

**Corresponding Author**

*E-mail: fyen@hit.edu.cn or fyen18@hotmail.com. Tel: +86-1343-058-9183.

**Author Contributions**

The manuscript was written through contributions of all authors. All authors have given approval to the final version of the manuscript. ‡These authors contributed equally.

REFERENCES

1. Yen, F.; Meng, L.; Gao, T.; Hu, S. Magnetic Ordering of Ammonium Cations in $NH_4I$, $NH_4Br$, and $NH_4Cl$. *J. Phys. Chem. C.* **2019**, *123*, 23655-23660.
2. Meng, L.; Peng, C. X.; He, C.; Yen, F. Magnetic Properties of $NH_4Al(SO_4)_2 \cdot 12H_2O$: Evidence of Glass Behavior Based on Proton Orbitals. *Scr. Mater.* **2021**, *205*, 114184.
3. Kozlenko, D. P.; Lewicki, S.; Wasicki, J.; Kozak, A.; Nawrocik, W.; Savenko, B. N. Ammonium ion dynamics in $NH_4I$ at high pressure. *Mol. Phys*. **2001**, *99*, 427-433.
4. Meng, L.; He, C.; Yen, F. Magnetoelectric Coupling Based on Protons in Ammonium Sulfate. *J. Phys. Chem. C.* **2020**, *124*: 17255-17261.
5. Samantaray, R.; Clark, R. J.; Choi, E. S.; Zhou, H.; Dalal, N. S. $M_{3-x}(NH_4)_xCrO_8$ (M= Na, K, Rb, Cs): A New Family of $Cr^{5+}$-Based Magnetic Ferroelectrics. *J. Am. Chem. Soc.* **2011**, *133*, 3792-3795.
6. Brüning, D.; Fröhlich, T.; Langenbach, M.; Leich, T.; Meven, M.; Becker, P.; Bohaty, L.; Gruninger, M.; Braden, M.; Lorenz, T. Magnetoelectric Coupling in the Mixed Erythrosiderite $[(NH_4)_{1-x}K_x]_2[FeCl_5(H_2O)]$, *Phys. Rev. B.* **2020**, *102*, 054413.
7. Sánchez-Andújar, M.; Gómez-Aguirre, L. C.; Pato Doldán, B.; Yáñez-Vilar, S.; Artiaga, R.; Llamas-Saiz, A. L.; Manna, R. S.; Schnelle, F.; Lang, M.; Ritter, F.; Haghighirad, A. A.; Señarís-Rodríguez, M. A. First-order Structural Transition in the Multiferroic Perovskite-like Formate $[(CH_3)_2NH_2][Mn(HCOO)_3]$. *CrystEngComm* **2014**, *16*, 3558-3566.
8. Liu, B.; Shang, R.; Hu, K. L.; Wang, Z. M.; Gao, S. A New Series of Chiral Metal Formate Frameworks of $[HONH_3][M^{II}(HCOO)_3]$ (M= Mn, Co, Ni, Zn, and Mg): Synthesis, Structures, and Properties. *Inorg. Chem.* **2012**, *51*, 13363-13372.
9. Gao, Z. R.; Sun, X. F.; Wu, Y. Y.; Wu, Y. Z.; Cai, H. L.; Wu, X. S. Ferroelectricity of the Orthorhombic and Tetragonal $MAPbBr_3$ Single Crystal. *J. Phys. Chem. Lett.* **2019**, *10*, 2522-2527.
10. Levy, H. A.; Peterson, S. W. Neutron Diffraction Determination of the Crystal Structure of Ammonium Bromide in Four Phases. *J. Amer. Chem. Soc.* **1953**, *75*, 1536-1542.
11. Gutowsky, H. S.; Pake, G. E.; Bersohn, R. Structural Investigations by Means of Nuclear Magnetism. III. Ammonium Halides. *J. Chem. Phys.* **1954**, *22*, 643-650.

12. Tsang, T.; Farrar, T. C.; Rush, J. J. Proton Magnetic Resonance and Hindered Rotation in Phosphonium Halides and Ammonium Iodide. *J. Chem. Phys.* **1968**, *49*, 4403-4406.
13. Pintar, M.; Sharp, A. R.; Vrscaj, S. Coexistence of Cubic and Tetragonal Magnetizations in Solid $NH_4I$ by Proton Spin-lattice Relaxation. *Phys. Lett. A.* **1968**, *27*, 169-170.
14. Xu, Y. Y.; Meng, L.; Zhao, M. M.; Peng, C. X.; Yen, F. Electric polarization and magnetic properties of $(NH_4)_{1-x}K_xI$ (x = 0.05–0.17). *J. Ally. Comp.* **2023**, *960*, 170685.
15. Berret, J. F.; Bostoen, C.; Hennion, B. Phase Diagram of the Dipolar Glass $K_{1-x}(NH_4)_xI$, *Phys. Rev. B.* **1992**, *46*, 747-750.
16. Murti, Y. V. G. S.; Prasad, P. S. Electrical Conductivity of $NH_4Cl$ and $NH_4Br$ Crystals in Phases I and II. *Physica B+C* **1975**, *79*, 243-255.
17. Momma, K.; Izumi, F. VESTA 3 for Three-Dimensional Visualization of Crystal, Volumetric and Morphology Data. *J. Appl. Crystallogr.* **2011**, *44*, 1272-1276.
18. Stephenson, C. C.; Landers, L. A.; Cole, A. G. Rotation of the Ammonium Ion in the High Temperature Phase of Ammonium Iodide. *J. Chem. Phys.* **1952**, *20*, 1044-1045.
19. Leung, P. S.; Taylor, T. I.; Havens Jr., W. W. Studies of Phase Transitions in Ammonium Salts and Barriers to Rotation of Ammonium Ions by Neutron Scattering Cross Sections as a Function of Temperature. *J. Chem. Phys.* **1968**, *48*, 4912-4918.
20. Bonilla, A.; Garland, C. W.; Schumaker, N. E. Low temperature X-ray investigation of $NH_4Br$. *Acta Cryst. A* **1970**, *26*, 156-158.
21. Goyal, P. S.; Dasannacharya, B. A. Neutron Scattering from Ammonium Salts. I. Coexisting Phases in $NH_4I$ and $NH_4^+$ Dynamics in Phase I and II. *J. Phys. C.* **1979**, *12*, 209.
22. Goyal, P. S.; Dasannacharya, B. A. Neutron Scattering from Ammonium Salts. II. Reorientational Motion of Ammonium Ions in Octahedral Environments. *J. Phys. C.* **1972**, *12*, 219.
23. Fehst, I.; Böhmer, R.; Ott, W.; Loidl, A.; Haussühl, S.; Bostoen, C. $(KI)_{1-x}(NH_4I)_x$: A dipolar glass. *Phys. Rev. Lett.* **1990**, *64*, 3139.
24. Funnel, N. P.; Bull, C. L.; Hull, S.; Ridley, C. Phase behaviour of ammonium bromide-$d_4$ under high pressure and low temperature; an average and local structure study. *J. Phys.: Condens. Matter* **2022**, *34***,** 325401.
25. Sorai, M.; Suga, H.; Seki, S. Phase transition in the ammonium bromide crystal: The thermal motion of the ammonium ion. *Bull. Chem. Soc. Jpn.* **1965**, *38*, 1125.
26. Seymour, R. T.; Pryor, A. W. Neutron Diffraction Study of $NH_4Br$ and $NH_4I$. *Acta Cryst. B.* **1970**, *26*, 1487-1491.
27. Zhang, W.; Xiong, R. G. Ferroelectric Metal–Organic Frameworks. *Chem. Rev.* **2012**, *112*, 1163-1195.
28. Meng, L.; He, C.; Ji, W.; Yen, F. Magnetic Properties of $NH_4H_2PO_4$ and $KH_2PO_4$: Emergence of Multiferroic Salts *J. Phys. Chem. Lett.* **2020**, *11*, 8297-8301.
29. Grinderslev, B.; Amdisen, M. B.; Skov, L. N.; Moller, K. T.; Kristensen, L. G.; Polanski, M.; Heer, M.; Jensen, T. R. New perspectives of functional metal borohydrides. *J. Ally. Comp.* **2022**, *896*, 163014.

**TOC graphic:**

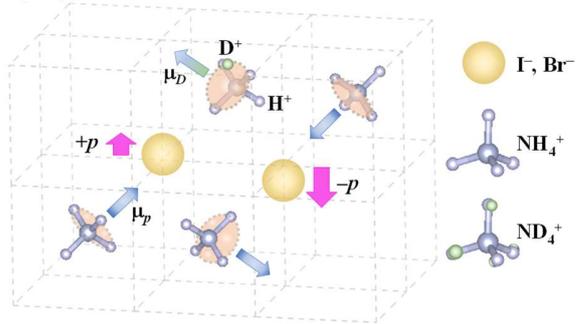